# Controlled Clustering of Superparamagnetic Nanoparticles using Block Copolymers : Design of New Contrast Agents for Magnetic Resonance Imaging

Jean-François Berret*, Nicolas Schonbeck, Florence Gazeau,
Delphine El Kharrat, Olivier Sandre, Annie Vacher and Marc Airiau

*Matière et Systèmes Complexes, UMR 7057 CNRS Université Denis Diderot Paris-VII, 140 rue de Lourmel F-75015 Paris France, Laboratoire Liquides Ioniques et Interfaces Chargées UMR 7612 CNRS Université Pierre et Marie Curie Paris-VI 4 place Jussieu, Boîte 51 F-75252 Paris Cedex 05 France, Rhodia, Centre de Recherches d'Aubervilliers, 52 rue de la Haie Coq, F-93308 Aubervilliers Cedex France*

RECEIVED DATE (automatically inserted by publisher); Author E-mail Address: jean-francois.berret@ccr.jussieu.fr

When polyelectrolyte-neutral block copolymers are mixed in aqueous solutions with oppositely charged species, stable complexes are found to form spontaneously. The mechanism is based on electrostatics, and on the compensation between the opposite charges. Electrostatic complexes exhibit a core-shell microstructure. In the core, the polyelectrolyte blocks and the oppositely charged species are tightly bound and form a dense coacervate microphase. The shell is made of the neutral chains and surrounds the core. In this paper, we report on the structural and magnetic properties of such complexes made from 6.3 nm diameter superparamagnetic nanoparticles (maghemite $\gamma$-$Fe_2O_3$) and cationic-neutral copolymers. The copolymers investigated are poly(trimethylammonium ethylacrylate methylsulfate)-*b*-poly(acrylamide), with molecular weights 5000-*b*-30000 g·mol$^{-1}$ and 110000-*b*-30000 g·mol$^{-1}$. The mixed copolymer-nanoparticle aggregates were characterized by a combination of light scattering and cryo-transmission electron microscopy. Their hydrodynamic diameters were found in the range 70 – 150 nm and their aggregation numbers (number of nanoparticles per aggregate) between several tens to several hundreds. In addition, Magnetic Resonance Spin-Echo measurements show that the complexes have a better contrast in Magnetic Resonance Imaging than single nanoparticles, and that these complexes could be used for biomedical applications.

## I - Introduction

Inorganic nanoparticles are currently used in a wide variety of material science and biomedical applications[1-3]. This is the case for magnetic[4-6] and luminescent nanocrystals[7-9] whose physical properties are exploited for imaging complex fluids at mesoscopic scales. The imaging techniques which make use of these particles are respectively Nuclear Magnetic Resonance and Imaging (NMR/MRI) and fluorescent microscopy. In biomedecine, however, one needs not only to detect or visualize the particles in cells, tissues or living organisms. The particles also have to be functionalized with macromolecules such as ligands, peptides and oligonucleotides to reach a target or to deliver a drug to a specific location[10]. For biomedical applications, then, inorganic particles are required to be concentrated or mixed in various environments. These transformations often result in the destabilization of the nanosol and in the aggregation/precipitation of the particles. When particles are irreversibly aggregated, they lose some important features, such as their large surface-to-volume ratio and their property to "get close" to biological entities[5]. In recent years, the colloidal stability of the inorganic and mineral nanoparticles has become a key issue in the control and design of novel nanostructures.

A broad range of techniques in chemistry and physical chemistry are now being developed for the stabilization of inorganic nanoparticles[3]. Among these methods are the adsorption of charged ligands or stabilizers on their surfaces[11-14], the layer-by-layer deposition of polyelectrolyte chains[15] and the surface-initiated polymerization (called "grafting from") resulting in high-density polymer brushes[16-18]. Other approaches have focused on the encapsulation of the particles in amphiphilic block copolymer micelles[19-21].

In the present paper, we are following an alternative strategy based on the principle of electrostatic complexation. The nanoparticles are complexed using asymmetric block copolymers, where one block is of opposite charge to that of the particles and the second block is neutral. Unlike amphiphilic block copolymers which are not soluble in water as unimers, charged-neutral diblocks self-assemble only in the presence of charged species. Recently, it has been shown that polyelectrolyte-neutral copolymers can associate in aqueous solutions with oppositely charged surfactants[22-27], macromolecules[1,28-31] and proteins[32,33] and are capable of building stable "supermicellar" aggregates with core-shell structures. The core (of radius ~ 10 – 20 nm) can be described as a dense coacervate microphase comprising the oppositely charged species. The corona (of thickness ~ 20 – 50 nm) is made by the neutral chains and ensures the colloidal stability of the whole. Typical molecular weights of the copolymers are comprised between 5 000 and 50 000 g·mol$^{-1}$.

In this paper, we show that it is possible to use electrostatic self-assembly in order to stabilize and associate $\gamma$-$Fe_2O_3$ superparamagnetic nanometer-sized particles in a controlled manner. We study the complexation between negatively charged particles with cationic-neutral block copolymers. Light scattering, transmission electron microscopy (cryo-TEM) and magnetic resonance spin-echo measurements have been utilized to study the copolymer-particles hybrids. Thanks to the strong electronic contrast of the iron atoms, we are able to visualize the morphology of the mixed aggregates and to derive their inner structure. As anticipated, we show that the core of the mixed aggregates is made of densely packed nanoparticles. The aggregation numbers (*i.e.* the number of particles per magnetic core) can be estimated directly from the cryo-TEM pictures. These numbers range from several tens to several hundreds, depending on the block copolymer used. Magnetic resonance spin-echo measurements indicate a significant increase of the ratio between the transverse and longitudinal relaxivities, which usually tests the efficiency of contrast agents in MRI. This result suggests that the polymer-nanoparticle hybrids designed by this technique could be used as $T_2$-contrast agents for biomedical applications.

## II - Experimental

### II. 1 – Sample Characterization and Preparation
*Polyelectrolyte-Neutral Diblock Copolymer*
The polyelectrolye-neutral diblocks were synthesized by controlled radical polymerization according to MADIX technology[34]. The polyelectrolyte portion is a positively charged poly(trimethylammonium ethylacrylate) block, whereas the neutral portion is poly(acrylamide)[25,26]. The chemical formulae of the monomers are shown in Fig. 1. Two molecular weights have been investigated, corresponding to 19 (5 000 g·mol$^{-1}$) and 41 (11 000 g·mol$^{-1}$) monomers in the charged blocks respectively and 420 (30 000 g·mol$^{-1}$) for the neutral one. In the following, the two copolymers are abbreviated as PTEA(5K)-*b*-PAM(30K) and PTEA(11K)-*b*-PAM(30K). Static and dynamic light scattering were performed on dilute copolymer solutions to determine the molecular weight and hydrodynamic diameter of the chains. As for the molecular weights, the nominal and experimental values were found in good agreement[26]. The average hydrodynamic diameters from dynamic light scattering (value of the quadratic term in the cumulant analysis) are $D_H^{pol}$ = 13 nm for the PTEA(5K)-b-PAM(30K) and $D_H^{pol}$ = 11 nm for PTEA(11K)-b-PAM(30K). The differences between the two values can be explained by the slightly different polydispersity of the two polymers. Note finally that since poly(trimethylammonium ethylacrylate) is a strong polyelectrolyte, its ionization state does not depend on pH. All solutions were prepared with 18.2 MΩ ultrapure Milli-Q water at pH 8.

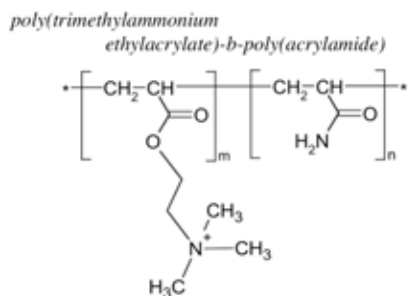

**Figure 1** : Chemical structure of the diblock copolymer PTEA-b-PAM investigated in the present work. The abbreviation PTEA stands for poly(trimethylammonium ethylacrylate methylsulfate) and PAM for poly(acrylamide).

*II.1.2 – Iron Oxide Nanoparticles*
Superparamagnetic nanoparticles of maghemite ($\gamma$-Fe$_2$O$_3$) were synthesized by alkaline co-precipitation of iron II and iron III salts[35,36] and sorted according to size by successive phase separations[37]. The nanosols were characterized using vibrating sample magnetometry, neutron and light scattering experiments. From the shape of the magnetization *versus* excitation curve, the size distribution of the particles was obtained[38]. This distribution is well described by a log-normal function with an average diameter of $\overline{D}^{nano}$ = 6.3 ± 0.1 nm and by a polydispersity of s = 0.23 ± 0.02, where s is defined as the ratio between the standard deviation and the first moment of the distribution. Using for the mass density the value of 5.1 g·cm$^{-3}$, the weight average molecular weight of the particles is $M_w^{nano}$ = 870 000 g·mol$^{-1}$ and the polydispersity index $M_w/M_n$ = 1.75. For the batch used, the magnetization at saturation was found to be lower than that of bulk maghemite ($M_S$ = 2.6·10$^5$ *versus* 4.2·10$^5$ A·m$^{-1}$). The reason for this decrease is attributed to the magnetic disorder of the iron spins at the surface of the particles[39]. At neutral pH (7 – 8), the particles are stabilized by electrostatic interactions thanks to the citrate ligands adsorbed by the particles[40,41]. ζ-potential measurements confirm that the citrate-coated nanoparticles are negatively charged (ζ = - 32.1 mV) and thus of opposite sign with respect to the polyelectrolyte PTEA block. The radius of gyration $R_G^{nano}$ and the hydrodynamic diameter $D_H^{nano}$ were estimated by neutron and dynamic light scattering techniques, respectively at $R_G^{nano}$ = 3.05 ± 0.06 nm and $D_H^{nano}$ = 11 nm. The small-angle neutron scattering experiments were performed on the PACE spectrometer at the Laboratoire Léon Brillouin (CEA-Saclay, France).

*II.1.3 - Sample Preparation*
Polymer-nanoparticle complexes are obtained by mixing pure solutions prepared at the same concentration and pH. The relative amount of each component is monitored by the mixing ratio X, which is defined as the volume of iron oxide sol relative to that of the polymer. Here, X is preferred to the charge ratio between the two species, essentially because the structural charges borne by the particles are not accurately known. With these notations and with the values of the molecular weights provided above, X = 1 corresponds to a solution with 20 polymers per particle. The weight concentrations in iron oxide and polymers in the mixed solutions are $c_{pol}$ = c/(1+X) and $c_{nano}$ = Xc/(1+X), where X = 0 (resp. 1) denotes the pure polymer (resp. particle) solution. For the mixed solutions, no macroscopic phase separation or precipitation was observed during or after the mixing. For the copolymer with the largest polyelectrolyte block, PTEA(11K)-b-PAM(30K), submicronic aggregates were observed and found to sediment after several days due to the large difference between the iron oxide density (5.1 g·cm$^{-3}$) and that of the suspending liquid.

### II. 2 - Experimental Techniques
Static and dynamic light scattering are performed on a Malvern / Amtec Macrotron spectrometer for the measurements of the Rayleigh ratio $R_\theta(\mathbf{c})$ and of the collective diffusion constant D(**c**). The light source used is a He/Ne laser operating at an incident power of 20 mW and at a wavelength λ = 633 nm. Light scattering is used to determine the preferred mixing ratio between the nanoparticles and the polymers. To this aim, the Rayleigh ratio was measured for mixed solutions with X ranging from X = 0.01 to X = 100 at a scattering angle of θ = 90°. With the spectrometer operating in dynamical mode, the collective diffusion coefficient D(**c**) was determined at θ = 90° and in the range c = 0.01 wt. % – 1 wt. %. From the value of D(c) extrapolated at c = 0, the hydrodynamic diameter of the colloids was calculated according to the Stokes-Einstein relation, $D_H$ = $k_BT/3\pi\eta_0 D(c\rightarrow 0)$, where $k_B$ is the Boltzmann constant, T the temperature (T = 298 K) and $\eta_0$ the solvent viscosity ($\eta_0$ = 0.89·10$^{-3}$ Pa·s). The autocorrelation functions of the scattered light were fitted using both cumulant and CONTIN analysis, both approaches yielding consistent values for $D_H$. Because of the absorption of the incident light by the iron oxide sols, the transmittance at 633 nm was measured separately by UV-visible spectrometry as a function of the iron molar concentration and the scattered intensities were corrected accordingly.

Cryo-transmission electron microscopy (cryo-TEM) experiments were performed on polymer-nanoparticle solutions made at the preferred mixing ration $X_P$ = 1 and concentration c = 0.2 wt. %. For cryo-TEM experiments, a drop of the solution is put on a TEM-grid covered by a 100 nm-thick polymer perforated membrane. The drop is blotted with filter paper and the grid is quenched rapidly in liquid ethane in order to avoid the crystallization of the aqueous phase. The membrane is finally transferred into the vacuum column of a TEM-microscope (JEOL 1200 EX operating at 120 kV) where it is maintained at liquid nitrogen temperature. The magnification for the cryo-TEM experiments was selected at 40 000×.

Magnetic resonance relaxometry has been used to measure the longitudinal ($T_1$) and transverse ($T_2$) proton relaxation times for the solutions described previously. The measurements were performed using a Brucker Minispec PC120 spectrometer operating at a magnetic field of 0.47 T and at a temperature T = 37 °C. This magnetic field value corresponds to a proton Larmor frequency of 20 MHz. Longitudinal and transverse relaxation times were obtained from inversion-recovery and from Carr-Purcell Meiboom Gill spin echo pulse sequences respectively, in agreement with procedures described in Ref.[42]. The accuracy in the relaxation time determination is ± 5 %.

## III – Results and Discussion
### III. 1 – Static and Dynamic Light scattering

Fig. 2 displays the Rayleigh ratio $R_{\theta=90°}$ for PTEA(5K)-*b*-PAM(30K) and PTEA(11K)-*b*-PAM(30K) with X ranging from $10^{-2}$ to 100 (c = 0.2 wt. %, T = 25° C). For the two polymers, the scattered intensity as function of X is found to pass through a sharp maximum around X = 1. For all values of X (with the exception of X = 0 and X = ∞ which describe the pure solutions), the data for the 11 K-polyelectrolyte block are above those obtained with the 5 K-system.

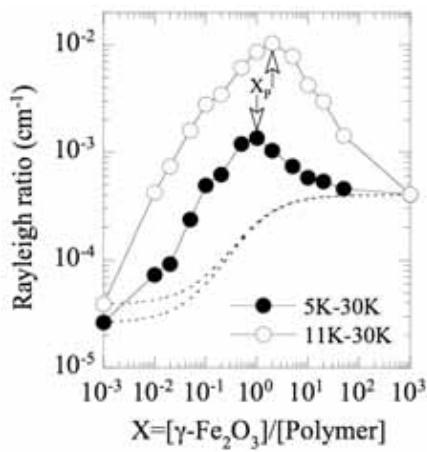

**Figure 2 :** Rayleigh ratio $R_\theta(X)$ measured by static light scattering for mixed copolymer- nanoparticles aggregates as function of the mixing ratio X. The intensities were collected at scattering angle θ = 90° using two different copolymers, PTEA(5K)-*b*-PAM(30K) (closed circles) and PTEA(11K)-*b*-PAM(30K) (empty circles) at concentration c = 0.2 wt. % and temperature T = 25° C. The dotted lines represent the scattering intensities calculated assuming that the two components remain unassociated. The arrows indicate the preferred mixing ratios $X_P$.

By definition, for disperse macromolecules and colloids, the scattered intensity extrapolated at zero concentration and zero scattering angle is proportional to the product $M_{w,app} c$, where $M_{w,app}$ is the apparent weight-averaged molecular weight of the scattering entities and c their weight concentration[43]. According to this hypothesis, the scattering intensity of solutions in which particles and polymers would remain unassociated can be computed. These calculations are displayed in Fig. 2 as dotted lines: they are far below the experimental data by one to two orders of magnitude. The excess scattering evidenced in Fig. 2 is interpreted as arising from large scale aggregates resulting from the association between nanoparticles and copolymers. We assume here that although the data are not taken at zero concentration and zero scattering angle, they are indicative of the actual variation of the product $M_{w,app} c$ as a function of $X^{44,45}$.

In electrostatic self-assembly, it has been found repeatedly that there exists a mixing ratio ($X_P$) at which all the components present in the solution form complexes[24,29,46]. This is the ratio where the number density of complexes is greatest, *i.e.* where the scattering intensity presents a maximum as a function of X. From the X-dependence of the Rayleigh ratio shown in Fig. 2, we found $X_P = 1$ for aggregates made with PTEA(5K)-*b*-PAM(30K) and $X_P = 2$ for those obtained with PTEA(11K)-*b*-PAM(30K). Using the molecular weights of the single components, the average numbers of polymers per particle at $X_P$ is estimated[45]. We found $\bar{n}^{pol}/\bar{n}^{nano} = 25$ for clusters made with PTEA(5K)-*b*-PAM(30K) and $\bar{n}^{pol}/\bar{n}^{nano} = 11$ for those made with PTEA(11K)-*b*-PAM(30K). Expressed in terms of charges, *i.e.* of charges borne by the polyelectrolytes, these numbers correspond in both cases to ~ 450 positive elementary charges.

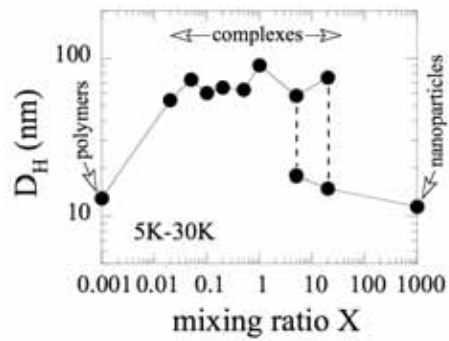

**Figure 3 :** Hydrodynamic diameter $D_H$ as a function of the mixing ratio X for mixed solutions made from PTEA(5K)-*b*-PAM(30K) block copolymers and iron oxide nanoparticles. For large values of X, a second diffusive mode associated to single nanoparticles is indicated. The hydrodynamic diameter $D_H$ for the individual components are $D_H^{pol} = 13$ nm and $D_H^{nano} = 11$ nm.

Dynamic light scattering performed on PTEA(5K)-*b*-PAM(30K)/iron oxide mixed solutions at c = 0.2 wt. % reveals the presence of purely diffusive relaxation modes for all values of X. Fig. 3 displays the evolution of the average hydrodynamic diameter $D_H$ derived from the time dependence of the autocorrelation functions and from the Stokes-Einstein relationship. For X > 0.01, $D_H$ ranges between 60 nm and 80 nm. At large X (X > 5), a second mode associated to the single nanoparticle became apparent. In this range, the autocorrelation is fitted by a double exponential decay. $D_H$-values as large as 60 – 80 nm are well above those of the individual components of the system. Again, this suggests the formation of mixed polymer-nanoparticle aggregates. The polydispersity index resulting from the cumulant analysis is found within a range of 0.10 to 0.25. These values are slightly above those found in similar studies using proteins[32,33] or surfactant[26].

## III.2 – Cryo-TEM

Fig. 4 displays a picture obtained by cryo-TEM of the iron oxide nanosol (c = 0.2 wt. %, X = ∞, no added polymer). The field shown on the figure is ~ 100×100 nm$^2$ and was obtained with a 40 000× magnification. This magnification allowed us to clearly distinguish γ-Fe$_2$O$_3$ nanoparticles, with diameters ranging from 5 to 10 nm. An image analysis on 472 particles provides an accurate size distribution (Fig. 4b). It is found to be described by a log-normal function with an average diameter $\bar{D}^{nano} = 6.3 \pm 0.2$ nm and a polydispersity s = 0.28 ± 0.04. These findings are in agreement with the magnetization results noted earlier : $\bar{D}^{nano} = 6.3 \pm 0.1$ nm and s = 0.23 ± 0.02.

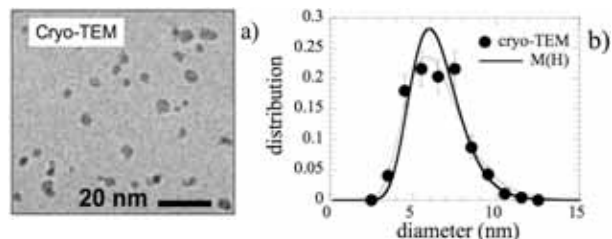

**Figure 4 :** a) Cryo-TEM picture from an iron oxide nanoparticle solution at c = 0.2 wt. %. b) Size distribution obtained by an image analysis performed on 472 citrate-coated particles. The data points are fitted using a log-normal function (continuous line, thin) with an average diameter $\bar{D}^{nano} = 6.3 \pm 0.2$ nm and a polydispersity s = 0.28 ± 0.04. These data are compared to the distribution received from vibrating sample magnetometry, noted M(H) and shown as a thick continuous line.

Cryo-transmission electron microscopy was performed on mixed polymer-nanoparticle solutions and the results illustrated in Figs. 5 to 6. Only mixed solutions prepared at the preferred mixing ratios were studied with this technique (c = 0.2 wt. %). The results are displayed in Fig. 5 for a sample made from PTEA(5K)-*b*-PAM(30K) diblocks and in Fig. 6 for one made from PTEA(11K)-*b*-PAM(30K). The photographs on the left hand side cover spatial fields that are approximately 1×1 µm². These dimensions demonstrate that the solutions contain well dispersed clusters of nanoparticles, results which are consistent with the visual observations of the solutions and with the light scattering data. In Fig. 5, a closer inspection reveals that the aggregates are slightly anisotropic, with sizes between 20 and 50 nm. The anisotropy is more pronounced with PTEA(11K)-*b*-PAM(30K) copolymers. For these copolymers, the aggregates are polydisperse in size and morphology, with maximum dimensions ranging from 100 nm to 200 nm. The fields of observation around two aggregates in Fig. 5 (aggregates **a** and **b**) and two in Fig. 6 (aggregates **c** and **d**) have been enlarged in order to underline the microstructure of the clusters (insets). For the cluster **a** highlighted in Fig. 5, six visible nanoparticles are assembled into a daisy-like structure. For the aggregates **b**, **c** and **d**, the morphology is more complex and it involves several tens to several hundreds of nanoparticles.

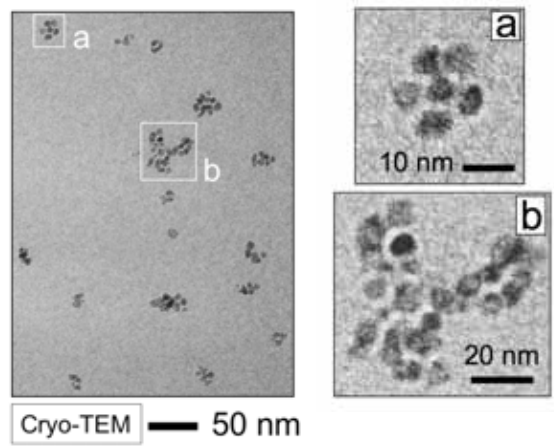

**Figure 5 :** Cryo-TEM image of mixed aggregates obtained by complexation of PTEA(5K)-*b*-PAM(30K) and iron oxide nanoparticles. The total concentration is c = 0.2 wt. % and X = $X_P$ (=1). Insets : A zoom of the fields of observation around aggregates **a** and **b** enabled us to dicern the densely packed 6.3 nm diameter nanoparticles inside the aggregates.

The average interparticle distance in the aggregates has been estimated by image analysis of 72 nanoparticle pairs. It was found to be 8.1 ± 0.1 nm, *i.e.* slightly larger than the diameter of a particle ($\overline{D}^{nano}$ = 6.3 nm). The value of 8.1 nm corresponds to a magnetic volume fraction of 0.32 ± 0.03 in the clusters. This findings are similar to the micellar volume fractions (0.4 – 0.5) observed for mixed surfactant-copolymer aggregates[25]. From the value of the particle volume fraction, we can derive the distribution of aggregation numbers, as well as the number and weight average aggregation numbers, $\overline{N}_n$ and $\overline{N}_w$ respectively. We found $\overline{N}_w$ = 32 for complexes made with the 5K block copolymers (and a polydispersity of 2.2) and $\overline{N}_w$ ~ 150 for those made with the 11K block copolymers (Table I). For this second system, the polydispersity could not be estimated owing to the small number of aggregates examined. In our forthcoming publication, the distributions and their derivation will be described in detail.

Due to the low electronic contrast between the copolymers and water[27], the presence of organic compounds in and around the aggregates can not be directly inferred from the cryo-TEM pictures. The presence of a polymer brush surrounding the clusters and made from poly(acrylamide) blocks can be deduced however from the comparison between electron microscopy and dynamic light scattering experiments carried out on the same solutions. For the sample using the shorter polyelectrolyte block (PTEA(5K)-*b*-PAM(30K) in Fig. 5), an image analysis was performed. It was assumed here that the clusters can be correctly described by ellipsoids of revolution, and that their projection can be accounted for by an ellipse with a major and a minor axis, noted *a* and *b* respectively. This approximation was found to hold for the majority of the aggregates. As a result of this analysis, the distribution functions for the minor and major axis were obtained and the equivalent hydrodynamic diameter was determined for this population of clusters, using the expression $\overline{D}_{eq}^{TEM} = a/G(\rho)$ [47]. Here, $\rho$ is the axial ratio *b*/*a* and $G(\rho) = (1-\rho^2)^{-1/2} \ln\{(1+(1-\rho^2)^{1/2})/\rho\}$, a function characteristic of prolate aggregates[47]. We found $\overline{D}_{eq}^{TEM}$ = 40 ± 5 nm, a value that is well below the actual hydrodynamic diameter measured by light scattering for the very same solution, $D_H$ = 70 ± 10 nm (see Table I).

| | $D_H$ (nm) | $\overline{N}_w$ | $R_1$ (s⁻¹·mM⁻¹) | $R_2$ (s⁻¹·mM⁻¹) | $R_2/R_1$ |
|---|---|---|---|---|---|
| block copolymers $M_w$ = 5K – 30 K $M_w$ = 11K – 30 K | 13 11 | -- | 0 | 0 | -- |
| γ-Fe₂O₃-nanoparticle | 11 | 1 | 23.5 ± 0.5 | 39 ± 2 | 1.7 |
| copolymer-nanoparticle aggregates made with : | | | | | |
| PTEA(5K)-*b*-PAM(30K) | 70 ± 10 | 32 (2.2) | 24 ± 1 | 74 ± 4 | 3.1 |
| PTEA(11K)-*b*-PAM(30K) | 170 ± 20 | ~ 150 | 17.5 ± 0.5 | 162 ± 4 | 9.3 |

**Table I :** Hydrodynamic diameter ($D_H$), weight average aggregation number ($\overline{N}_w$), longitudinal ($R_1$) and transverse ($R_2$) relaxivities and relaxivity ratio ($R_2/R_1$) characterizing the polymers, the nanoparticles and the polymer-nanoparticle mixed aggregates. $\overline{N}_w$ denotes the number of magnetic nanoparticles per aggregate, determined from an image analysis of 187 clusters. The number in parenthesis in the fourth line is the polydispersity index for the clusters.

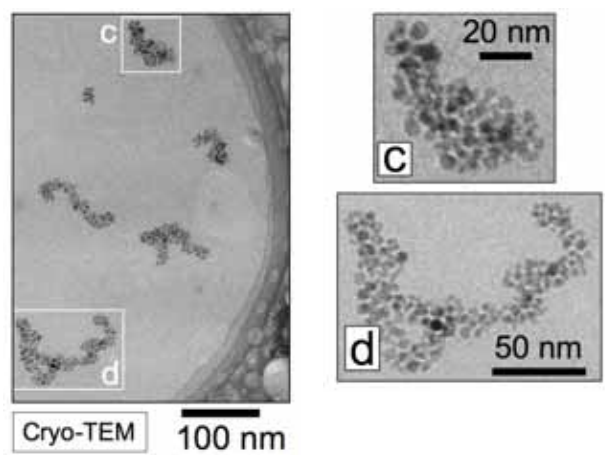

**Figure 6 :** Same experimental conditions as in Figure 5 for hybrids obtained from PTEA(11K)-*b*-PAM(30K) copolymers. Insets : A zoom of the fields of observation around aggregates **c** and **d**.

This suggests that the nanoparticle clusters displayed in Figs. 5 and 6 are surrounded by a polymer corona made from the neutral blocks of poly(acrylamide). It is worth mentioning that the thickness of the corona derived here is of the order of 10 - 20 nm, and therefore agrees with the value of the corona thickness determined using the surfactant-copolymer mixed systems[26].

### III.3 – Relaxometry

Figs. 7a and 7b show the inverse relaxation times $1/T_1$ and $1/T_2$ as a function of the iron molar concentration [Fe] for the citrate-coated nanoparticles and for the nanoparticle-copolymer aggregates. Samples were prepared at $X_P = 1$ for PTEA(5K)-*b*-PAM(30K) and at $X_P = 2$ for PTEA(11K)-*b*-PAM(30K), both at concentration c = 0.2 wt. %. In order to monitor the inversion-recovery and spin echo pulse sequences, the solutions were diluted by a factor of 10 to 1000. For these experiments, light scattering was used to verify that the size and microstructure of the polymer-nanoparticle complexes were not modified under dilution. In Figs. 7a and 7b, the inverse relaxation times were found to vary linearly with the iron concentration, according to the following equation[48]:

$$\frac{1}{T_{1,2}([Fe])} = R_{1,2} \cdot [Fe] + \frac{1}{T_{1,2}^0} \qquad (1)$$

where $R_1$ and $R_2$ are the longitudinal and transverse relaxivities, respectively. The intercepts $1/T_{1,2}^0$ are the proton inverse relaxation times in pure water. At a Larmor frequency of 20 MHz, we found $T_1^0 = 4.0$ s and $T_2^0 = 1.6$ s. The data in Figs. 7a and 7b shows that with increasing cluster size, the longitudinal relaxivity $R_1$ exhibits a slight decrease, but remains around 20 s$^{-1}$·mM$^{-1}$. For clusters such as those in Figs. 5 and 6, $R_1$ is actually close to that of the bare particles. The transverse relaxivity $R_2$ on the contrary increases noticeably. $R_2$ starts at $39 \pm 2$ s$^{-1}$·mM$^{-1}$ for the bare nanoparticles, rises up to $74 \pm 4$ for clusters prepared with the 5K-30K copolymer, and culminates at $162 \pm 4$ for clusters made with the 11K-30K chains. This data is shown in Table I, together with the hydrodynamic diameters of the particles and hybrids. Such values for $R_1$ and $R_2$ suggest that the polymer-nanoparticle aggregates created by electrostatic self-assembly could serve as $T_2$ contrast agents, especially consideration is given to the magnetic core being protected by a neutral polymer corona.

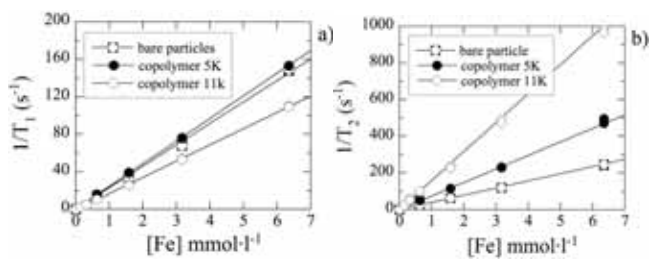

**Figure 7 :** Inverse longitudinal and transverse relaxation times $1/T_1$ (a) and $1/T_2$ (b) for mixed copolymer-particle hybrids as a function of the iron molar concentration [Fe]. The solutions were prepared at $X = X_P$ and c = 0.2 wt. % and were diluted in the range [Fe] = 1 – 10 mM (corresponding to $c_{nano}$ = 0.1 – 0.01 wt. %) for the inversion-recovery and spin echo pulse sequences. The straight lines were calculated according to Eq. 1. The inverse relaxation times for the pure ferrofluid solution are also included for comparison.

It is interesting to compare these results to those obtained for other magnetic clusters and assemblies. For submicronic unilamellar vesicles loaded with ferrofluid, Billotey et al.[49] and Martina et al.[42] have shown the same variation of the relaxivity ratio, $R_2/R_1$ with increasing iron loading. However, for these systems the relaxivities vary in the opposite way : $R_1$ decreases with iron loading, whereas $R_2$ increases barely above the value of the single nanoparticles[42].

Recently, Manuel-Perez et al. have developed magnetic nanosensors for detecting proteins, DNA and enzymatic activities for *in vitro* diagnostic experiments[50]. For 50 nm-magnetic coated particles, clustering was achieved by the addition of complementary oligonucleotides or peptide sequences. These changes in the solution structure were accompanied by an enhancement of the relaxation rate $1/T_2$[50]. More recently, Roch and coworkers have used calcium counterions to progressively destabilize a commercial magnetic nanosol[51]. The transverse relaxivity was monitored during destabilization, and again the transverse relaxivity was found to increase noticeably. Finally, Ai et al. have obtained similar results for magnetic nanoparticles encapsulated into the hydrophobic cores of 20 - 100 nm polymeric micelles[20]. The results quoted in the last three studies support those reported in the present work. They all indicate that clusters of magnetic nanoparticles have enhanced transverse relaxivities $R_2$, and relaxivity ratios $R_2/R_1$.

### IV – Conclusion

In this paper, we investigated the formation of new hybrid aggregates comprising charged copolymers and nanoparticles. The mechanism at the origin of the self-assembly is based on electrostatic adsorption and charge compensation between oppositely charged species. The systems put under scrutiny were cationic-neutral hydrosoluble block copolymers and 6.3 nm diameter superparamagnetic $\gamma$-Fe$_2$O$_3$ nanoparticles.

The first section of the paper addresses the formation and the characterization of the microstructures resulting from this association. Two copolymers differing by polyelectrolyte block lengths (5 000 and 11 0000 g·mol$^{-1}$, respectively) and with the same neutral segment (poly(acrylamide)) of molecular weight 30 000 g·mol$^{-1}$) have been used for comparison. In this study, we have shown that the formation of the hybrid aggregates is spontaneous and reproducible, and that the aggregates exhibit long-term colloidal stability. Our first goal was to derive the preferred mixing ratio $X_P$, which corresponds to a state where the charges borne by the chains and by the particles approximately compensate. We found $X_P = 1$ for PTEA(5K)-*b*-PAM(30K) and $X_P = 2$ for PTEA(11K)-*b*-PAM(30K).

The inner structure of the aggregates is elucidated by combining dynamic light scattering and transmission electron microscopy performed on samples that were quenched down to nitrogen temperature (cryo-TEM). Owing to the good electronic contrast of the iron oxide nanoparticles, it was possible to directly analyze aggregates resulting from electrostatic complexation, something that, to the best of our knowledge, had not been accomplished previously. The nano-hybrids are made of a core comprising tens to hundreds densely packed magnetic nanoparticles, as well as a surrounding corona made from the neutral poly(acrylamide) blocks. We found that the magnetic clusters are larger with PTEA(11K)-*b*-PAM(30K) than with PTEA(5K)-*b*-PAM(30K). These findings suggest that the molecular weight of the polyelectrolyte blocks is an important control parameter for the determination of the size of the hybrids.

From these experiments, analogies have been made between the present system and previously investigated systems, with an emphasis placed on copolymer-surfactant complexes[25-27]. The comparison of the two systems, specifically regarding their core-shell structure, is striking. One noticeable difference however is the polydispersity in size or aggregation number of the mixed aggregates. This polydispersity is much larger with the mineral particles (2.2 *versus* 1.2 for the surfactant[27]) and it probably originates from the broader distribution of the $\gamma$-Fe$_2$O$_3$ nanocrystals. The growth of aggregates giving rise to elongated structures, such as those of Fig. 6 could also be responsible for the broad size distribution.

The sequel of the paper deals with the contrast properties of the mixed polymer-particle aggregates in magnetic resonance imaging,

and their possible benefits for biomedical applications. A significant result is that the transverse relaxivity, $R_2$ is noticeably increased with the size of the magnetic clusters. The ratio $R_2/R_1$, which is usually an important parameter in estimating the efficiency between $T_2$-contrast agents, increases from 1.7 for the citrate-coated nanoparticles[52] to 9.3 for the clusters built with PTEA(11K)-*b*-PAM(30K) copolymers. Owing to the simplicity of the complexation phenomenon and to the fact that the magnetic clusters are thereafter protected by a diffuse neutral polymer corona, the present technique might be promising for the building of stable nanostructures devoted to biomedical applications.

**Acknowledgements** : We thank Frederic Gohran and Yoann Lalatonne for their assistance during the experiments, Jean-Paul Fortin, Jean-Claude Bacri, Valérie Cabuil for fruitful discussions. Mathias Destarac from Rhodia (Centre de Recherches d'Aubervilliers, France) is acknowledged for providing us with the polymers. We are grateful to Fabrice Cousin and to the Laboratoire Léon Brillouin (CEA-Saclay, France) for the neutron experiments on the uncoated particles (experiment #7660). We also thank Olivier Clément from Hopital Necker (Paris, France) for the use of the Bruker Minispec relaxometer. This research is supported by the Centre National de la Recherche Scientifique.

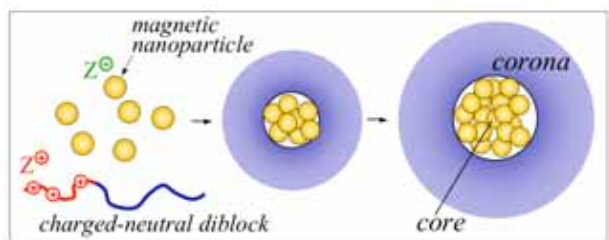
TOC Figure